\documentclass[aps,prb,twocolumn]{revtex4} 
\usepackage{times} 
\usepackage{graphics,dcolumn} 
\usepackage{amsmath,amssymb} 
\usepackage{graphicx}  
\usepackage{color}

\begin{document}

\title{Application of the anisotropic bond model to 
second-harmonic generation from amorphous media} 
\author{E.~J.~Adles, D.~E.~Aspnes} \affiliation{Department of Physics, 
NC State University, Raleigh, NC 27695-8202} 

\date{\today} 
\begin{abstract}
	As a step toward analyzing second-harmonic generation 
(SHG) from crystalline Si nanospheres in glass, we develop an 
anisotropic bond model (ABM) that expresses SHG in terms of physically 
meaningful parameters and provides a detailed understanding of the 
basic physics of SHG on the atomic scale. Nonlinear-optical (NLO) 
responses are calculated classically via the four fundamental steps of 
optics:  evaluate the local field at a given bond site, solve the 
force equation for the acceleration of the charge, calculate the 
resulting radiation, then superpose the radiation from all charges.  
Because the emerging NLO signals are orders of magnitude weaker and 
occur at wavelengths different from that of the pump beam, these steps 
are independent.  Paradoxically, the treatment of NLO is therefore 
simpler than that of linear optics (LO), where these calculations must 
be done self-consistently.  The ABM goes beyond previous bond models 
by including the complete set of underlying contributions:  
retardation (RD), spatial-dispersion (SD), and magnetic (MG) effects, 
in addition to the anharmonic restoring force acting on the bond 
charge.  Transverse as well as longitudinal motion is also considered. 
We apply the ABM to obtain analytic expressions for SHG from amorphous 
materials under Gaussian-beam excitation.  These materials represent 
an interesting test case not only because they are ubiquitous but also 
because the anharmonic-force contribution that dominates the SHG 
response of crystalline materials and ordered interfaces vanishes by 
symmetry. The remaining contributions, and hence the SHG signals, are 
functions entirely of the LO response and beam geometry, so the only 
new information available is the anisotropy of the LO response at the 
bond level. The RD, SD, and MG contributions are all of the same order 
of magnitude, so none can be ignored.  Diffraction is important not 
only in determining the pattern of the emerging beam but also the 
phases and amplitudes of the different terms.  The plane-wave 
expansion that gives rise to electric quadrupole/magnetic dipole 
effects in LO appears as RD here.  Using the paraxial-ray 
approximation, we reduce the results to the isotropic case in two 
limits, that where the linear restoring force dominates (glasses), and 
that where it is absent (metals).  Both forward- and backscattering 
geometries are discussed.  Estimated signal strengths and conversion 
efficiencies for fused silica appear to be in general agreement with 
data, where available. Predictions are made that allow additional 
critical tests of these results.

\end{abstract}


\maketitle

\section{Introduction\label{sec-introduction}} Second-harmonic 
generation (SHG) is becoming an increasingly important diagnostic tool 
for a wide range of applications. It is a particularly important probe 
for studying planar interfaces between centrosymmetric crystals and 
overlayers with randomly directed bonds, because it is dipole-allowed 
only at the interface where the bonds are simultaneously asymmetric 
and well ordered. 

Recently, Figliozzi et al.~\cite{Figliozzi2005} found that SHG signals 
generated in transmission from crystalline Si nanospheres (nSi) 
dispersed in glass were enhanced significantly when driven by two 
beams with crossed polarizations. Enhancement of any nonlinear-optical 
(NLO) signal is automatically of interest, because in principle NLO 
signals contain significantly more information about materials systems 
than the linear-optical (LO) response, yet are intrinsically much 
weaker. SHG from the dispersed-nSi configuration was recently analyzed 
from the macroscopic perspective by Brudny et al.~\cite{Brudny2000} 
and Moch\'{a}n et al.,~\cite{Mochan2003} in the former case for a 
single isolated nanosphere and the latter for arrays of nanospheres. 
These authors used the ``dipolium'' approximation,~\cite{Mendoza1997} 
where the inclusions and host are described macroscopically by linear, 
isotropic dielectric functions. The far-field SHG response was 
obtained by calculating the effective dipole of the inclusions as a 
spherical-harmonic expansion of the internal and external fields of a 
given nanosphere, then applying standard radiation equations. Various 
observations were explained, for example the $(\vec{E} \cdot \nabla 
)\vec{E}$ symmetry of the SHG intensity, its dependence on sphere 
size, the importance of screening in determining the contributions 
from the interiors of the nanospheres, the emission of SHG radiation 
in a cone for disordered dispersions of nanospheres, and the 
relatively small intensity of the SHG signal from glass.

While macroscopic treatments efficiently distinguish between allowed 
and forbidden contributions, they are unable to relate allowed 
responses to atomic-scale parameters, or to provide the same level of 
understanding of the different contributing processes. In particular, 
the following questions still need to be answered: (1) how does the 
SHG intensity from the nSi inclusions compare to that from planar 
Si--SiO$_2$ interfaces; (2) what are the relevant parameters; (3) what 
is the maximum intensity that can be obtained; and (4) is this maximum 
signal useful, or simply given by a combination of already known 
parameters? While much larger SHG signals might be expected from 
dispersions of nSi inclusions in a transmission configuration simply 
because the interface area greatly exceeds that of a planar interface, 
the larger area is offset by the fact that the first-order 
anharmonic SHG signals from the opposite sides of the nanospheres 
cancel. Therefore, the signal is proportional to the 
gradient of the driving field instead of the field 
itself.~\cite{Brudny2000} In addition, contributions are limited in 
depth to the coherence length in the material. Finally, there is the 
question of whether the enhanced SHG signals observed with dual-beam 
excitation provide useful information. The atomic-scale modeling 
done below shows that the contributions of the three underlying 
mechanisms, retardation (RD), spatial dispersion (SD), and magnetic 
(MG), can all be predicted from the LO response and beam 
characteristics, hence do not necessarily provide new information even 
though improved geometries may generate large signals. 

In addressing these issues we found it necessary to extend our 
previous simplified bond-hyperpolarizability model (SBHM), which 
expands on the even simpler isotropic force model discussed for 
example in Shen.~\cite{Shen2003} In the SBHM, SHG is expressed as 
radiation arising from the anharmonic motion of charge localized in 
bonds assuming that the only motion relevant to SHG is that along the 
bond direction itself. The SBHM successfully describes, with many 
fewer parameters than previously required, a wide range of NLO 
phenomena including SHG \cite{Powell2002} and FHG \cite{Hansen2003} 
from Si--insulator interfaces, dipole-forbidden SHG \cite{Peng2005a} 
and THG \cite{Peng2004} from centrosymmetric materials, and the 
generation of THz radiation from III--V semiconductor 
surfaces.~\cite{Peng2005} In addition, the parameters are physically 
meaningful, and by incorporating crystal symmetry at the atomic level, 
macroscopic tensor properties are obtained automatically. However, as 
recently, and correctly, noted by McGilp,~\cite{McGilp2007yq} the SBHM 
has limitations regarding quantitative interpretation. Given the 
simplicity of the approach this is not surprising, but it needs to be 
explored further. This is a second objective of this work.

Accordingly, in the present paper we generalize the SBHM to a more 
complete description, the anisotropic bond model (ABM), which includes 
charge motion transverse to the bond, RD, SD, and MG effects, 
including SD effects arising from beam geometry, and SHG signals for 
off-axis observation angles, i.e., the role of diffraction. In 
developing our expressions we follow the approach of Peng et al.,\cite{Peng2005a} 
framing the calculations in terms of the fundamental 4-step process of 
optics: (1) evaluate the local field at any given charge site that 
results from the driving (source) field; (2) solve the mechanical 
equation $\vec{F} = m\vec{a}$ to obtain the acceleration of the 
charge; (3) calculate the radiation that results from the 
acceleration; and (4) superpose the radiation from all contributing 
charges. For random media we show that step (4) factors into two 
parts: (4a) average over all possible bond orientations at a single 
site, then (4b) calculate the properties of the emerging beam by 
Fourier-transforming the envelope function of the incident radiation. 
Although not required here, if appreciable energy were transferred 
from the driving to the generated beams, then it would be necessary to 
(5) evaluate the energy extracted from the pump beam as a function of 
position, with the subsequent correction of the local fields evaluated 
in step (1). We find that for random materials the RD, SD, and MG 
contributions are all of the same order of magnitude and must all be 
considered.  Finally, all aspects, including off-axis observation and 
diffraction, combine to yield a much richer SHG response than 
previously assumed.

Aside from including bond anisotropy and additional mechanisms, the 
approach is essentially the NLO equivalent of that which Ewald 
\cite{Ewald1912} and Oseen \cite{Oseen1915} used nearly a century ago 
to derive the Ewald-Oseen theorem of LO. Paradoxically, from this 
perspective NLO is simpler than LO. In LO the radiated fields have 
the same wavelength as the driving field and similar intensities, so 
steps (1), (3), and (4) must be evaluated self-consistently. In 
contrast, for NLO the radiated fields are typically orders of 
magnitude weaker than the driving field and occur at different 
wavelengths, so all steps are effectively independent. This allows NLO 
calculations to be done sequentially, to levels of approximation that 
are also independent and may be adjusted to meet particular 
requirements. This is one of the few cases where a nonlinear problem 
is simpler than its linear equivalent.

Advantages of an atomic-scale formulation include a better 
understanding of the physics involved. In this classical model NLO is 
a result of distortions of the nominally sinusoidal waveform of the 
emitted radiation reaching the observer. The obvious contributing 
factor is anharmonic motion of a charge. This can be due to an 
anharmonic restoring force (intrinsic anharmonicity), spatial 
nonuniformity of the driving field (spatial dispersion), or the 
magnetic field associated with the driving wave. With respect to 
acceleration, there can clearly be no distinction between anharmonic 
motion resulting from an anharmonic restoring force, a field that is 
slightly larger at one limit of the excursion than the other, or a 
force that is velocity-dependent. All these effects enter in step (2). 
However, another source of distortion is the finite speed of light. 
This causes the signal reaching the observer from the far limit of the 
excursion to be delayed slightly relative to that from the near limit, 
resulting in a waveform distortion equivalent to phase- or 
frequency-modulation. The retardation contribution enters in step (3). 
Retardation involves the same first-order expansion of a plane-wave 
factor that leads to the electric quadrupole/magnetic dipole 
contribution of LO, but the physics is quite different.  This 
mathematical similarity has led to confusion in the past, and we 
clarify the distinction below.

Taking into account the complete set of mechanisms contributing to SHG 
became a larger project than expected, so in the present paper we 
restrict applications to single-beam excitation of disordered 
materials and reduction of the resulting expressions to the case where 
the bond charges are isotropically polarizable in LO.  We discuss two 
limits, first where the restoring force dominates the acceleration 
term (glasses), and second where the restoring force is absent 
(metals).  This reduction, done in the paraxial-ray approximation, 
highlights the roles of the different underlying mechanisms, the 
difference between forward- and backscattering configurations, and 
allows a simple expression for signal strength and conversion 
efficiency to be obtained.  The present work represents a necessary 
first step toward our goal of understanding, at the atomic level, SHG 
from Si nanoinclusions in glass under crossed-beam excitation, and is 
further justified by the fact that disordered materials are ubiquitous 
in many fields.

\section{Amorphous materials\label{sec-amorphous}}

\subsection{Fields at bond sites\label{ssec-bond-fields}}

In this section we consider step (1), define basic quantities, and 
discuss the connection between first- and second-harmonic fields. We 
suppose that the relevant quantities are electrons of charge $q = -e$ 
located in bonds $j$ at positions $\vec{r}_q = \vec{r}_j + \Delta 
\vec{r}_j$, where the $\vec{r}_j$ are the equilibrium positions of the 
charges relative to the origin of a coordinate system defined in the 
laboratory and the $\Delta \vec{r}_j$ are the displacements that 
result from the time-dependent forces acting on them. We represent the 
directions of the bonds by $\hat{b}_j$, where for Si--O bonds the 
$\hat{b}_j$ point from Si to O. 

For amorphous materials that are homogeneous on meso and macroscopic 
length scales, the driving field can be assumed to be approximately a 
plane wave with frequency $\omega$, envelope function $\vec{E}_o 
(\vec{r}_q )$, and wave vector $\vec{k}_o$ = $(\omega n_1 / 
c)\hat{k}_o$, where $n_1$ is the refractive index of the material at 
$\omega$. We assume that the Fresnel reflectance coefficients have 
been appropriately taken into account at the surface of the material 
to yield the correct amplitude $E_o$ of 
$\vec{E}_o(\vec{r}_q)$ within the medium. Then the field at the 
$j^{th}$ charge can be written to first order in $\Delta \vec{r}_j$ as 
\begin{equation}
	\begin{split}
		\vec{E} (\vec{r}_q, t) & = \vec{E}_o (\vec{r}_q \, ) 
e^{i\vec{k}_o \cdot \vec{r}_q - i \omega t } \\
		& = \vec{E}_o (\vec{r}_j + \Delta \vec{r}_j) 
e^{i\vec{k}_o \cdot (\vec{r}_j + \Delta \vec{r}_j ) - i \omega t } \\
		& \approx [ 1 + \Delta \vec{r}_j \cdot \nabla 
_{\vec{r}_j} ] \vec{E}_o (\vec{r}_j) e^{i\vec{k}_o \cdot \vec{r}_j - 
i\omega t} \\
		& = [ 1 + \Delta \vec{r}_j \cdot \nabla _{\vec{r}_j} ] 
\vec{E}_j e^{- i\omega t}. 
	\end{split}
	\label{ff1} 
\end{equation}
For clarity in the following equations, we let $\vec{E}_j$ = 
$\vec{E}_o (\vec{r}_j) e^{i \vec{k}_o \cdot \vec{r}_j}$ contain the 
spatial dependences of the envelope and phase. The SHG nature of the 
correction term follows because $\Delta \vec{r}_j$ is also 
proportional to $\vec{E}_j$, as shown below, so the gradient term 
nominally has a time dependence $e^{-i2 \omega t}$.

However, the coefficient of a $e^{-i2\omega t}$ term is not simply the 
product of the coefficients of the parent $e^{-i\omega t}$ terms, but 
must be reduced by a factor of 2 for the following reason. Observables 
are real quantities, so $e^{-i \omega t} = \cos( \omega t) + i \sin 
(\omega t)$ is actually shorthand for $Re ( e^{-i \omega t } ) = \cos 
( \omega t)$. Thus the product of two $e^{-i\omega t }$ terms is 
really a product $ \cos^2( \omega t)$, $\sin( \omega t) \cos ( \omega 
t)$, or $ \sin^2( \omega t)$, or some combination depending on the 
phases of the parent coefficients. All trigonometric identities taking 
$\omega t$ products into $2\omega t$ forms involve a factor of 
($1/2$). We introduce this factor in the far-field radiation 
expression Eq.~\eqref{Eff3}. We retain the $e^{- i 2 \omega t}$ 
notation so average intensities can be calculated in the usual way.

\subsection{Force equation\label{ssec-force-eq}}

The general form of the force equation for SHG is 
\begin{equation}
	\begin{split}
		\vec{F} = m \vec{a} & = m \frac {d^2 \Delta \vec{r} 
(t) } {dt^2} \\
		& = q \vec{E} ( \vec{r}, t) + q \frac{\vec{v}}{c} 
\times \vec{B}(\vec{r},t) \\
		& \quad \quad - \tilde{ \kappa}_1 \cdot \Delta \vec{r} 
(t) - \tilde{\kappa}_2 \cdot \cdot \Delta \vec{r} (t) \Delta \vec{r} 
(t) , 
	\end{split}
	\label{f1} 
\end{equation}
where $m$ is the mass of $q$ and $\tilde{\kappa}_{1}$ and 
$\tilde{\kappa}_{2}$ are second- and third-rank tensors describing the 
linear (Hooke's Law) and first-order anharmonic restoring forces, 
respectively, $\vec{v} = d \Delta \vec{r}/ dt$, and the magnetic-flux 
density $\vec{B} (\vec{r}_q,t)$ associated with the driving field is 
$\vec{B} (\vec{r}_{q},t) = -(ic/\omega) \nabla_{\vec{r}_{q}} \times 
\vec{E}(\vec{r}_{q},t)$. In contrast to the SBHM, we do not assume the 
force equation to be one-dimensional. To find the displacements 
$\Delta \vec{r}_j$, we substitute Eq.~\eqref{ff1} into the force 
equation to obtain 
\begin{equation}
	\begin{split}
		m \frac{d^2 \Delta \vec{r}_j (t) } {d t^2} = q & \left 
[ 1 + \Delta \vec{r}_j (t) \cdot \nabla_{\vec{r}_j} \right ] \vec{E}_j 
e^{-i\omega t} \\
		& + \frac{q}{c} \frac{d \Delta \vec{r}_j (t) }{dt} 
\times \vec{B}_j e^{-i\omega t} \\
		& - \tilde{\kappa}_1 \cdot \Delta \vec{r}_j(t) - 
\tilde{\kappa}_2 \cdot \cdot \Delta \vec{r}_j (t) \Delta \vec{r}_j(t) 
, 
	\end{split}
	\label{f2} 
\end{equation}
where $\vec{ B }_j = (-i c / \omega) \nabla_{\vec{r}_j} \times 
\vec{E}_{\vec{r}_j}$. From the form of $\vec{E}(\vec{r}_q ,t) $, we 
can assume that 
\begin{equation}
	\Delta \vec{r}_j (t) = \Delta \vec{r}_{1j} e^{-i\omega t} + 
\Delta \vec{r}_{2j} e^{-i2\omega t} , \label{drr} 
\end{equation}
where $\Delta \vec{r}_{1j}$ and $\Delta \vec {r}_{2j}$ are time 
independent. Substituting this expression in Eq.~\eqref{f2} yields 
\begin{equation}
	\begin{split}
		& -m \omega^{2} \Delta \vec{r}_{1j} e^{-i\omega t} - 4 
m \omega^2 \Delta \vec{r}_{2j} e^{-i 2 \omega t} \\
		& \quad \quad = q \left ( 1+ ( \Delta \vec{r}_{1j} 
e^{-i\omega t} ) \cdot \nabla_{\vec{r}_j} \right ) \vec{E}_j e^{- 
i\omega t} \\
		& \quad \quad \quad - q (\Delta \vec{r}_{1j} e^{- 
i\omega t} ) \times (\nabla_{r_j} \times \vec{E}_{j} e^{-i\omega t}) 
\\
		& \quad \quad \quad - \tilde{\kappa}_1 \cdot \left ( 
\Delta \vec{r}_{1j} e^{-i\omega t} + \Delta \vec{r}_{2j} e^{-i2\omega 
t} \right ) \\
		& \quad \quad \quad - \tilde{\kappa}_2 \cdot \cdot 
\Delta \vec{r}_{1j} \Delta \vec{r}_{1j} e^{-i2\omega t} . \label{f3} 
	\end{split}
\end{equation}
Since $\Delta\vec{r}_j$ is at least first-order in $\vec{E}$, the 
magnetic term is at least second-order in $\vec{E}$. Since 
we are only concerned with SHG, we neglect terms of order $(3 \omega)$ 
and $(4 \omega)$, which would contribute to THG \cite{Peng2004} and 
FHG \cite{Hansen2003} respectively.

Isolating the first-harmonic terms we have
\begin{equation}
	-m \omega^2 \Delta \vec{r}_{1j} = q \vec{E}_j - 
\tilde{\kappa}_1 \cdot \Delta \vec{r}_{1j} . \label{firstord} 
\end{equation}
While this can be solved in general by matrix methods, we now 
introduce the approximation that $\tilde{\kappa}_1$ and 
$\tilde{\kappa}_2$ are diagonal in the local coordinate system of the 
bond, where the $z$ axis is defined by the unit vector $\hat{b}$ 
parallel to the bond. Diagonalization is equivalent to assuming that 
the bonds are rotationally symmetric. While bonds in some 
systems are not rotationally symmetric, we make this simplifying 
assumption to elucidate the underlying physics. Obviously, if desired 
all tensor components of the restoring forces could be kept.

We also define the unit vector $\hat{t}$, which is perpendicular to 
$\hat{b}$ and lies in the $\hat{b}-\vec{E}$ plane. Thus $\hat{t}$ is 
given by 
\begin{equation}
	\hat{t} = \left (\vec{E} - \hat{b} ( \hat{b} \cdot \vec{E} ) 
\right ) / \sqrt{ \vec{E} ^2 - \left ( \hat{b} \cdot \vec{E} \right ) 
^2 }. \label{t-hat} 
\end{equation}
Then $\tilde{\kappa}_1$ and $\tilde{\kappa}_2$ can be written as 
\begin{equation}
	\tilde{\kappa}_1 = \hat{b} \hat{b} \kappa_{1l} + \hat{t} 
\hat{t} \kappa_{1t} , \label{k1lt} 
\end{equation}
\begin{equation}
	\tilde{\kappa}_2 = \hat{b} \hat{b} \hat{b} \kappa_{2l} , 
\label{k2l} 
\end{equation}
where $\kappa_{1l}$ and $\kappa_{2l}$ are the longitudinal linear and 
anharmonic restoring-force coefficients, respectively, and 
$\kappa_{1t}$ is that for transverse displacements. With the 
assumption of rotational symmetry, $\kappa_{2t}$ does not exist. 
However, transverse contributions are still possible through the RD, 
SD, and MG terms. 

Substituting these expressions into Eq.~\eqref{f3} and taking dot 
products with $\hat{b}$ and $\hat{t}$ leads to the two first-order 
equations 
\begin{equation}
	\Delta \vec{r}_{1jl} = \Delta r_{1jl} \hat{b}_j = \frac {q ( 
\hat{b}_j \cdot \vec{E}_j ) } { \kappa_{1l} - m \omega^2 } \hat{b}_j ; 
\label{dr1l} 
\end{equation}
\begin{equation}
	\Delta \vec{r}_{1jt} = \Delta r_{1jt} \hat{t}_j = \frac {q ( 
\hat{t}_j \cdot \vec{E}_j ) } {\kappa_{1t} - m \omega^2 } \hat{t}_j . 
\label{dr1t} 
\end{equation}
Repeating the process for the second-order terms leads to
\begin{equation}
	\begin{split}
		\Delta \vec{r}_{2jl} & = \Delta r_{2jl} \hat{b}_j \\
		& = \left[ q ( \Delta \vec{r}_{1j} \cdot 
\nabla_{\vec{r}_j} ) ( \hat{b}_j \cdot \vec{E}_j ) \right. \\
		& \quad - q ( \Delta \vec{r}_{1j} \times ( 
\nabla_{\vec{r}_{j}} \times \vec{E}_{j} )) \cdot \hat{b}_j \\
		& \quad - \kappa_{2l} \Delta r_{1jl} \Delta r_{1jl} 
\Big] \hat{b}_j / (\kappa_{1l} - 4 m \omega^2) ; 
	\label{dr2l} 
	\end{split}
\end{equation}
\begin{equation}
	\begin{split}
		\Delta \vec{r}_{2jt} & = \Delta r_{2jt} \hat{t}_j \\
		& = \left[ q ( \Delta \vec{r}_{1j} \cdot 
\nabla_{\vec{r}_j} ) ( \hat{t}_j \cdot \vec{E}_j ) \right. \\
		& \quad - q ( \Delta \vec{r}_{1j} \times ( 
\nabla_{\vec{r}_{j}} \times \vec{E}_{j} )) \cdot \hat{t}_j \Big] 
\hat{t}_j / (\kappa_{1t} - 4 m \omega^2) ; 
	\label{dr2t} 
	\end{split}
\end{equation}
\begin{equation}
	\begin{split}
		\Delta \vec{r}_{2j(b \times t)} & = \Delta r_{2j(t 
\times b)} (\hat{b}_j \times \hat{t}_j) \\
		& - q \left[ ( \Delta \vec{r}_{1j} \times ( 
\nabla_{\vec{r}_{j}} \times \vec{E}_{j} )) \cdot (\hat{b}_j \times 
\hat{t}_j ) \right] \\
		& \quad \times (\hat{b}_j \times \hat{t}_j )/ 
(\kappa_{1t} - 4 m \omega^2) .
	\label{dr2bt} 
	\end{split}
\end{equation}
Equation~\eqref{dr2bt} is necessary because the magnetic force 
generates a component that is perpendicular to both $\hat{b}$ and 
$\hat{t}$. Equation~\eqref{dr2l} shows that there is no qualitative 
distinction between the intrinsic anharmonicity of a bond and an 
anharmonicity generated by a field, as expected. These are the 
expressions from which the acceleration, and therefore the far-field 
signal, will be calculated.

\subsection{Far-field radiation from accelerated charges
\label{far-field}}

We now consider step (3). We follow the development of Peng et 
al.,~\cite{Peng2005a} but take into account explicitly the reduction 
in propagation speed caused by refractive indices $n_\nu$ that are 
different from 1. The two that need to be considered are $n( \omega ) 
 = \sqrt{\epsilon (\omega)}  = n_1$ for the incoming wave 
and $n( 2\omega ) = \sqrt{\epsilon (2 \omega)}  = n_2$ 
for the emitted SHG radiation. Accordingly, we write $\vec{k}_o = k_o 
\hat{k}_o = (\omega n_1 / c) \hat{k}_o$ and $\vec{k} = k \hat{k} = (2 
\omega n_2 /c) \hat{k}$ for the incident and emerging radiation, 
respectively, where $\hat{k}$ points in the direction of the observer.

The general expression for the four-potential of an accelerated point 
charge in the medium in Fourier-component form is 
\begin{equation}
	\begin{split}
		&\left [ \phi \left ( \vec{r}, t \right ) , \vec{A} 
\left ( \vec{r},t \right ) \right ]_\nu = \frac {1} {c} \int d^3 r' 
dt' \\
		& \quad \quad \quad \quad \times \left [ c \rho \left 
( \vec{r} \,', t' \right ) , \vec{J} \left ( \vec{r} \,', t' \right ) 
\right ]_\nu G_\nu (\vec{r},\vec{r} \,',t,t'), 
	\end{split}
	\label{fa} 
\end{equation}
where 
\begin{equation}
	G_\nu (\vec{r},\vec{r}\, ',t,t') = \frac { \delta \left ( t - 
t' - \frac {n_{\nu}} {c} \left | \vec{r} - {\vec{r}}\ ' \right | 
\right ) } {\left | \vec{r} - \vec{r} \, ' \right | } \label{GF} 
\end{equation}
is the Green function, $\rho$ and $\vec{J}$ = $\rho\vec{v}$ are the 
charge and current densities, respectively, and $n_{\nu}$ is $n_1$ or 
$n_2$ according to whether the frequency of interest is $\omega$ or 
$2 \omega$. Here, $\rho$ and $\vec{J}$ are associated with the 
$j^{th}$ point charge $q$ located at $\vec{r}_q = \vec{r}_{j} + \Delta 
\vec{r}_j(t) $. Then 
\begin{equation}
	\rho_j \left ( \vec{r} \,', t' \right ) = q \delta \left ( 
\vec{r} \,' - \vec{r}_j - \Delta \vec{r}_j (t') \right ) ; \label{rho} 
\end{equation}
\begin{equation}
	\begin{split}
		\vec{J}_j \left ( \vec{r} \,', t' \right ) & = \rho_j 
\left ( \vec{r} \,', t' \right ) \frac {d \Delta \vec{r}_j (t')} 
{dt'}\\
		& = \left ( q \frac {d \Delta \vec{r}_j (t')} {dt'} 
\right ) \delta \left ( \vec{r} \,' - \vec{r}_j - \Delta \vec{r}_j 
(t') \right ) . \label{rj} 
	\end{split}
\end{equation}
The far-field $\vec{E}^{ff}_j(\vec{r},t) $ that results 
from $q$ is given by 
\begin{equation}
	\begin{split}
		\vec{E}^{ff}_j ( \vec{r}, t) & = - \frac {1} {c} \frac 
{ 
		\partial \vec{A}_j ( \vec{r},t ) } { 
		\partial t} - \nabla \phi_j ( \vec{r},t ) \\
		& = - \frac {1} {c} \frac { 
		\partial \vec{A}_{j\perp} (\vec{r},t ) } { 
		\partial t} 
	\end{split}
	\label{Eff1} 
\end{equation}
where $ \vec{A}_{j\perp} (\vec{r},t)$ is the component of 
$\vec{A}_j(\vec{r},t)$ that is perpendicular to the line between the 
origin $\vec{r}_j + \Delta \vec{r}_j $ of the radiation and the 
observer at $\vec{r}$. The second line of Eq.~\eqref{Eff1} follows 
because $\nabla \phi_j$ in the first line removes the longitudinal 
component of $\vec{A}_j$, leaving a purely transverse potential. Thus 
we need evaluate only $\vec{A}_j(\vec{r},t)$. This can be obtained 
relatively simply because $\vec{A}_j$ is already of first order in 
$\vec{v}/c$, where $\vec{v}$ is the velocity of $q$. 

Substituting Eq.~\eqref{rj} into Eq.~\eqref{fa} and performing the 
integration over $\vec{r} \,'$ yields 
\begin{equation}
	\begin{split}
		&\vec{A}_j (\vec{r},t) \\
		& \quad = \frac {q} {c} \int dt' \frac {d \Delta 
\vec{r}_j (t')} {dt'} \frac { \delta \left ( t - t' - \frac {n_{\nu}} 
{c} | \vec{r} - \vec{r}_j - \Delta \vec{r}_j (t') | \right ) } { | 
\vec{r} - \vec{r}_j - \Delta \vec{r}_j (t') | }. \label{a2} 
	\end{split}
\end{equation}
The integration over $t'$ is nontrivial because $\Delta \vec{r}_j (t 
\,' ) $ is also a function of $t'$. However, to first order in $1/c$ 
we can expand 
\begin{equation}
	\frac {n_\nu} {c} \left | \vec{r} - \vec{r}_j - \Delta 
\vec{r}_j (t') \right | \approx \frac {n_\nu r} {c} - \frac {n_\nu} 
{c} \hat{k} \cdot \vec{r}_j - \frac {n_\nu} {c} \hat{k} \cdot \Delta 
\vec{r}_j (t') , \label{expansion} 
\end{equation}
so 
\begin{equation}
	\begin{split}
		& \delta \left ( t - t' - \frac {n_\nu} {c} | \vec{r} 
- \vec{r}_j - \Delta \vec{r}_j (t') | \right ) \\
		& \approx \delta \left ( t_o - t' + \frac {n_\nu} {c} 
\hat{k} \cdot \Delta \vec{r}_j (t') \right ) , 
	\end{split}
	\label{t2} 
\end{equation}
where $t_o = t - n_\nu r/c + n_\nu \hat{k} \cdot \vec{r}_j / c\,$. This 
is still a self-consistent expression, but to first order in $1/c$ we 
can substitute $t_o$ for $t'$ in the argument of $\Delta \vec{r}_j 
(t')$. We obtain finally 
\begin{equation}
	t' = t_{ret} \approx t_o + \frac {n_\nu} {c} \hat{k} \cdot 
\Delta \vec{r}_j \left ( t_o \right ) , \label{t3} 
\end{equation}
where $t_{ret}$ is the retarded time. The integral over $t'$ can now 
be performed, and we obtain 
\begin{equation}
	\vec{A}_j ( \vec{r},t) = \frac {q} {rc} \left ( \frac {d 
\Delta \vec{r}_j (t') } {dt'} \right ) _{t' = t_{ret} } . \label{a2.5} 
\end{equation}
Substituting Eq.~\eqref{drr} into Eq.~\eqref{a2.5} yields the 
contribution from the $j^{th}$ charge: 
\begin{equation}
	\begin{split}
		\vec{A}_j( \vec{r},t ) & = - \frac{i \omega q} {rc} 
\left ( \Delta \vec{r}_{1j} e^{-i \omega t' } + 2 \Delta \vec{r}_{2j} 
e^{ - i2 \omega t' } \right ) _{t' = t_{ret} } \\
		& = - \frac {i \omega q} {rc} \left ( \Delta 
\vec{r}_{1j} e^{- i \hat{k} k_o \cdot \Delta \vec{r}_1j } e^{ - ik_o 
\hat{k} \cdot \vec{r}_j} e^{ik_o r - i\omega t} \right. \\
		& \left. \quad \quad \quad \quad \quad \quad + 2 
\Delta \vec{r}_{2j} e ^ {- i \vec{k} \cdot \vec{r}_j } e ^ {i kr - i 2 
\omega t } \right) \\
		& = - \frac {i \omega q} {rc} \Delta \vec{r}_{1j} e^{- 
i \hat{k} k_o \cdot \vec{r}_j} e^{ik_or - i\omega t} \\
		& \quad \quad - \frac{ \omega ^2 q n_{2}}{rc^2} \Delta 
\vec{r}_{1j} ( \hat{k} \cdot \Delta \vec{r}_{1j} ) e^{- i \vec{k} 
\cdot \vec{r}_j } e^{ ikr - i2\omega t} \\
		& \quad \quad - \frac{i2 \omega q}{rc} \Delta 
\vec{r}_{2j} e^{- i\vec{k} \cdot \vec{r}_j } e^{ ikr - i2\omega t } . 
	\end{split}
	\label{a3} 
\end{equation}
The far field signal $\vec{E}^{ff}_j$ then follows from 
Eq.~\eqref{Eff1}:
\begin{equation}
	\begin{split}
		\vec{E}^{ff}_j ( \vec{r},t ) = & \left[ \tilde{I} - 
\hat{k}\hat{k} \right] \cdot \left[ \frac{ \omega ^2 q}{r c^2} \Delta 
\vec{r}_{1j} e^{-i{k}_o \hat{k} \cdot \vec{r}_j} e^{ik_or - i\omega t} 
\right. \\
		& - i\frac{ \omega^3 q n_{2} }{rc^3} \Delta 
\vec{r}_{1j} (\hat{k} \cdot \Delta \vec{r}_{1j} ) e^{-i\vec{k} \cdot 
\vec{r}_j} e^{ikr - i2\omega t} \\
		& \left. + \frac{2 \omega^2 q}{rc^2} \Delta 
\vec{r}_{2j} e^{-i \hat{k} \cdot \vec{r}_j} e^{ikr - i2\omega t} 
\right] .
	\end{split}
	\label{Eff3} 
\end{equation}
Here, $\tilde{I} - \hat{k}\hat{k}$ is the projection operator that 
eliminates the longitudinal component and hence performs the function 
of $-\nabla \phi _j$. As with $-\nabla \phi _j$, $\tilde{I} - 
\hat{k}\hat{k}$ does not affect the orthogonal component, which will 
be found to be significant when we discuss term cancellations in Secs. 
III B and III C. In the two nonlinear terms of Eq.~\eqref{Eff3}, we 
have now incorporated the factor of $(1/2)$ associated with the change 
of time dependence from $(e^{-i\omega t})^2$ to $e^{- i2\omega t}$ as 
discussed in Sec.~\ref{ssec-bond-fields}. 

Equation~\eqref{Eff3} is a general expression for linear and 
second-order far-field radiation from a moving charge in terms of 
displacements from its equilibrium position. The first term is the 
linear response. The second term is the RD contribution, and the third 
is a combination arising from the spatial dependence of the field (SD, 
MG) and the intrinsic anharmonicity of the bond (the third term in 
Eq.~\eqref{dr2l}). Because the RD contribution originates in 
propagation, not acceleration, the use of the common expression 
\begin{equation}
	\vec{E}^{ff} = - \frac{1}{c^2} \frac { 
	\partial ^2 \vec{a}_\perp }{ 
	\partial t^2} \label{common} 
\end{equation}
leads for this term to an error of a factor of 2.

To address a point that has caused difficulty in the past, we note 
that the RD term above and the electric quadrupole/magnetic dipole 
(EQ/MD) terms of LO both result from an expansion of a phase term 
$e^{i\vec{k} \cdot \vec{r} }$ to first order in $\vec{k} \cdot 
\vec{r}$. However, the physics, and consequently the nature of 
$\vec{E}^{ff}$, is different in the two situations. In LO 
$\rho(\vec{r},t)$ is assumed to be a moderately extended but 
stationary charge density with a multiplicative time dependence $e^{-
i\omega t}$, thus having the form $\rho(\vec{r},t) = \rho_o(\vec{r}) 
e^{-i\omega t}$. Here, the $t'$ integration is trivial but the 
$\vec{r}\,'$ integration is not. Performing the $t'$ integration 
yields a multiplicative time factor $e^{- i\omega t}$ and a phase term 
$e^{i\vec{k}\cdot\vec{r}\,'}$ that is part of the electrostatic Green 
function. Because the current $\vec{J}$ needed to calculate $\vec{A}$ 
has no obvious representation in this case, appropriate 
vector-calculus identities are used to convert the integration of 
$\vec{J}$ into a first-moments integration of 
$\vec{r}\,'\rho(\vec{r}\,')$.~\cite{Jackson1998fk} The dipole 
approximation follows by taking $e^{i\vec{k}\cdot\vec{r}\,'} = 1$, 
with higher-multipole moments generated from higher-order expansion 
terms.~\cite{Shen2003,Jackson1998fk}  Thus the LO expansion 
gives rise to multipole moments but no higher harmonics. 

In contrast, in the present work $\rho(\vec{r},t)$ describes a moving 
point charge $q\delta(\vec{r}\,' - \vec{r}_o(t))$, where $\vec{r}_o(t) 
= \vec{r}_j + \Delta\vec{r}e^{-i\omega t}$. As seen above, the 
$\vec{r}\,'$ integration is now trivial but the $t'$ integration is 
not. We obtain here higher harmonics but no multipole moments. Thus 
what Peng et al. labeled EQ/MD in ref.~\cite{Peng2005a} is due to 
retardation.  That in ref.~\cite{Brudny2000} is actually due to 
spatial dispersion. 

\subsection{Superposition of radiation; averaging and 
diffraction\label{ssec-superposition}}

In the following we assume that the charges are driven coherently, so 
fields must be added rather than intensities. This is expected, and 
the validity of the assumption demonstrated experimentally by the 
vanishing of SHG for amorphous materials in the forward direction. 

Returning to Eqs.~\eqref{ff1}, \eqref{dr1l}--\eqref{dr2t}, and 
\eqref{Eff3}, the $\vec{r}_j$ dependence of the SHG signal is either 
$E_o^2 ( \vec{r}_j ) e^{i (2 \vec{k}_o - \vec{k} ) \cdot \vec{r}_j}$ 
or a derivative of the form $E_o ( \vec{r}_j ) ( 
\partial E_o( \vec{r}_j ) / 
\partial x) e^{i (2 \vec{k}_o - \vec{k} ) \cdot \vec{r}_j}$. Both are 
slowly varying on the atomic scale, whereas the bond directions 
$\hat{b}$ and $\hat{t}$ vary essentially randomly from site to site. 
Given this large difference of scale we can factor step (4) into two 
parts: averaging over bond orientations, effectively at a single site; 
then evaluating the sum over all $\vec{r}_j$. 

\subsubsection{Bond averages\label{ssec-averages}}

We consider first averaging over bond directions. This is accomplished 
by writing 
\begin{equation}
\begin{split}
\hat{b} & = b_x \hat{x} + b_y \hat{y} + b_z \hat{z} \\
& = \hat{x} sin\theta cos\phi + \hat{y} sin\theta sin\phi + 
\hat{z} cos\theta 
\end{split}
\label{b}
\end{equation}
then performing the operation 
\begin{equation}
	\left<f (\hat{b},\hat{t}\,) \right> = \frac{1}{4\pi} \int 
d\Omega \, f(\hat{b},\hat{t}\,) 
	\label{avg} 
\end{equation}
The calculation is simplified by grouping the products involving bond 
directions into dyadics, triadics, etc., then considering symmetry. 
For example for LO the bond averages that need to be evaluated occur 
as dyadics $\hat{b}\hat{b}$ and $\hat{t}\hat{t}$. In the 
Cartesian-coordinate representation $\hat{b}\hat{b}$ has 9 terms 
$b_x b_x \hat{x} \hat{x}$, $b_x b_y \hat{x}\hat{y}$, etc., but only 3 
survive the averaging process because any component involving an odd 
number of projections averages to zero. 

By this reasoning the triadic $\hat{b}\hat{b}\hat{b}$ clearly vanishes 
identically, so by Eq.~\eqref{dr2l} there can be no $\kappa_{2l}$ 
contribution to SHG in amorphous materials. Not surprisingly, both 
microscopic and macroscopic considerations therefore lead to the same 
conclusion. However, the implications here go further. The terms that 
remain are functions only of the LO response and the configuration 
geometry, so the amount of new information obtainable by SHG in 
amorphous materials is limited to the separation of longitudinal and 
transverse components of the LO response, no matter what geometries 
are used to enhance the SHG signal.

We now consider the RD contribution to SHG. The terms that need to be 
considered are $\hat{b}\hat{b}\hat{b}\hat{b}$, 
$\hat{b}\hat{b}\hat{t}\hat{t}$ and permutations, and 
$\hat{t}\hat{t}\hat{t}\hat{t}$. If desired, the results can be 
decomposed into irreducible tensor representations, although we do not 
do this here. In the calculations that follow we take advantage of the 
absence of a preferred direction in amorphous materials. Hence without 
loss of generality we assume that $\vec{k}_o = k_o \hat{z}$ and 
$\vec{E}_o = E_o \hat{x}$. The expression to be evaluated is then
\begin{equation}
	\begin{split}
		\vec{E}^{ff}_{RD,j} &= -i \frac{ \omega ^3 q n_2 }{15 
rc^3} \left[ \tilde{I} - \hat{k} \hat{k} \right] \cdot \frac{1}{4\pi} 
\int d \Omega \, \left( \Delta \vec{r}_{1j} \cdot \hat{k} \right) \\
		& \quad \quad \quad \quad \quad \times \Delta 
\vec{r}_{1j} e^{- i \vec{k} \cdot \vec{r}_j } e^{ikr - i2\omega t}. 
	\end{split}
	\label{glassRD1} 
\end{equation}
The result is 
\begin{equation}
	\begin{split}
		\vec{E}^{ff}_{RD} & = - i \frac{ \omega^3 q n_2}{15 r 
c^3} \left[ \tilde{I} - \hat{k} \hat{k} \right] \cdot \\
		& \quad \quad \times \left[ \hat{x} (\hat{k} \cdot 
\hat{x}) ( 3 C_{l}^2 + 4 C_{l}C_{t} + 8 C_{t}^2) \right. \\
		& \left. \quad \quad \quad \quad \quad + \left(\hat{y} 
(\hat{k} \cdot \hat{y}) + \hat{z} (\hat{k} \cdot \hat{z}) \right) 
(C_{l} - C_{t})^2 \right] \\
		& \quad \quad \times \left[\sum_{\vec{r}_j} 
E_o^2(\vec{r}_j) e^{ i( 2 \vec{k}_o - \vec{k} ) \cdot \vec{r}_j 
}\right] e^{ikr - i2\omega t} , 
	\end{split}
	\label{glassRD} 
\end{equation}
where to simplify the expression we define 
\begin{equation}
	C_{l} = \frac{q }{\kappa_{1l} - m \omega^2} ; \quad C_{t} = 
\frac{q }{\kappa_{1t} - m \omega^2} . 
\end{equation}
We write the $x$ component etc. of $\vec{k}$ as 
$k(\hat{k}\cdot\hat{x})\hat{x}$ so we can move the magnitude of 
$\vec{k}$ to the prefactor and therefore eliminate an easily 
overlooked source of error. The separate longitudinal and transverse 
contributions can be obtained by setting $C_{t}$ = 0 or $C_{l}$ = 0, 
respectively. The sum over $\vec{r}_j$ is clearly a Fourier transform 
of the square of the envelope function of the driving beam, and will 
be evaluated in Sec.~\ref{ssec-diffraction}. 

We consider next the contributions from SD. These are given by 
\begin{equation}
	\begin{split}
		& \vec{E}^{ff}_{SD,j} = \frac{2 \omega ^2 q}{r c^2} \\
		& \quad \times \left[ \tilde{I} - \hat{k}\hat{k} 
\right] \cdot \frac {1}{4\pi} \int d\Omega \left[ \frac {q ( \Delta 
\vec{r}_{1j} \cdot \nabla_{\vec{r}_j} ) ( \hat{b} \cdot \vec{E}_j ) } 
{ \kappa_{1l} - 4m \omega^2 } \hat{b} \right. \\
		& \left. \quad \quad \quad \quad \quad \quad \quad 
\quad + \frac {q ( \Delta \vec{r}_{1j} \cdot \nabla_{\vec{r}_j} ) ( 
\hat{t} \cdot \vec{E}_j ) } { \kappa_{1t} - 4m \omega^2 } \hat{t} 
\right] \\
		& \quad \quad \quad \quad \quad \quad \times e^{-i 
\vec{k} \cdot \vec{r}_j} e^{ikr - i2\omega t }. 
	\end{split}
	\label{glassSD} 
\end{equation}
We evaluate Eq.~\eqref{glassSD} by dividing the field gradient into 
longitudinal and transverse parts with respect to $\vec{k}_o$, i.e., 
letting $\nabla_{ \vec{r}_{j} } E ( \vec{r}_j ,t) = \hat{x} \frac{ 
\partial E }{ \partial x} + \hat{y} \frac{ \partial E }{ \partial y} + 
i \hat{z} k_{o} $. After performing the averages we obtain
\begin{equation}
	\begin{split}
		&\vec{E}^{ff}_{SD}(\vec{r},t) = \frac{2 \omega ^2 q } 
{15 r c^2 } \left[ \tilde{I} - \hat{k} \hat{k} \right] \cdot 
\sum_{\vec{r}_j} E_o (\vec{r}_j )\\
		& \times \bigg\{ \hat{x} \frac{ 
		\partial E_o ( \vec{r}_j ) }{ 
		\partial x} \left( 3 C_{l} D_{l} + 2 C_{l} D_{t} + 2 
C_{t} D_{l} + 8 C_{t} D_{t} \right) \\
		& \left. \quad + \left[\hat{y} \frac{ 
		\partial E_o ( \vec{r}_j ) }{ 
		\partial y} + i \hat{z} k_o E_o(\vec{r}_j)\right] 
\left( C_{l} - C_{t} \right) \left( D_{l} - D_{t} \right) \right\}\\
		& \times e^{i(2\vec{k}_o - \vec{k}) \cdot \vec{r}_j } 
e^{ikr - i2 \omega t } 
	\end{split}
	\label{glassSD2} 
\end{equation}
where
\begin{equation}
	D_{l} = \frac{q }{\kappa_{1l} - 4 m \omega^2} ; \quad D_{t} = 
\frac{q }{\kappa_{1t} - 4 m \omega^2} . 
\end{equation}
The envelope function for the $z$ component is the same as that 
for the RD contribution, but those for $\hat{x}$ and $\hat{y}$ involve 
gradients of the driving field. 

The MG contribution is given by 
\begin{equation}
	\begin{split}
		& \vec{E}^{ff}_{MG,j} = - \frac{2 \omega^2 q}{r c^2} 
\left[ \tilde{I} - \hat{k}\hat{k} \right] \cdot \frac {1}{4\pi} \\
		& \quad \times \int d\Omega \left\{ D_{l} \left[ ( 
\Delta \vec{r}_{1j} \times (\nabla_{\vec{r}_j} \times \vec{E}_j )) 
\cdot \hat{b} \right] \hat{b} \right. \\
		& \quad \quad \quad \quad + D_{t} \left[ ( \Delta 
\vec{r}_{1j} \times (\nabla_{\vec{r}_j} \times \vec{E}_j )) \cdot 
\hat{t} \right] \hat{t}  \\
		& \left. \quad \quad \quad \quad + D_{t} \left[ ( 
\Delta \vec{r}_{1j} \times (\nabla_{\vec{r}_j} \times \vec{E}_j )) 
\cdot ( \hat{b} \times \hat{t}\,) \right] (\hat{b} \times \hat{t}\,) 
\right\} \\
		& \quad \quad \quad \quad \quad \quad \quad \quad 
\times e^{-i \vec{k} \cdot \vec{r}_j} e^{ikr - i2\omega t }. 
	\end{split}
	\label{glassMag} 
\end{equation}
Here, all three dimensions are involved.  By suitable vector-calculus 
identities the double-cross-product operation can be cast into 
apparent spatial-dispersion form, $(\vec{E} \cdot \nabla ) 
\vec{E}$,~\cite{Shen2003} but an exact cancellation of the resulting 
dominant terms makes this approach unproductive.  After performing the 
cross-product operations with the assumed propagation and field 
directions and then averaging over bond orientations we obtain 
\begin{equation}
	\begin{split}
		& \vec{E}^{ff}_{MG} = - \frac{2 \omega^2 q}{3 r c^2} 
\left[ \tilde{I} - \hat{k}\hat{k} \right] \cdot \sum_{\vec{r}_j} E_o 
(\vec{r}_j) \\
		& \quad \times \left( 2 C_{t} D_{l} + C_{l} D_{t} 
\right) \left[ \hat{y} \frac{ 
		\partial E_o (\vec{r}_j)}{ 
		\partial y} + i \hat{z} k_o E_o (\vec{r}_j) \right] \\
		& \quad \quad \quad \quad \quad \quad \quad \quad 
\times e^{i(2\vec{k}_o - \vec{k}) \cdot \vec{r}_j } e^{ikr - i2\omega 
t }. 
	\end{split}
	\label{glassMag2} 
\end{equation}

\subsubsection{Diffraction\label{ssec-diffraction}} 

We consider now the sums over $\vec{r}_j$. These not only yield the 
geometric properties of the emerging SHG beam, but also affect the 
phases and amplitudes of the prefactors of the individual 
constituents. In the derivation below we assume forward scattering, 
but will discuss backscattering in Sec.~\ref{back}. We consider 
throughout only single-beam excitation. Crossed-beam configurations 
follow the same principles but are complicated by the need to consider 
large observation angles, so will be treated in a subsequent paper. 

We assume that the incident beam is Gaussian. For the RD, $\hat{z}$ 
SD, and $\hat{z}$ MG contributions, the relevant sum is
\begin{equation}
	\sum_{ \vec{r}_j } E_o ^2 e^{-2(x^2 + y^2 )/W^2} e^{i (2 
\vec{k}_o - \vec{k} ) \cdot \vec{r}_j }, 
\end{equation}
where $W$ is the width of the incident beam and for our configuration 
$\vec{k}_o = k_o \hat{z}$. Converting the sum to an integral we have 
\begin{equation}
	\sum_{\vec{r}_j} \rightarrow N \int_{- \infty}^{\infty} dx dy 
\int_0^L dz , 
\end{equation}
where $N$ is the volume density and $L$ the thickness of the sample. 
The integrals are all standard and we find 
\begin{equation}
	\sum_{\vec{r}_j} E_o (\vec{r}_j)^2 e^{i ( 2\vec{k}_o - \vec{k} 
) \cdot \vec{r}_j} = i \frac{ \pi N W^2 E_o^2 }{2(2 k_o - k_z) } e^{- 
( k_x^2 + k_y^2 ) W^2 / 8}. \label{diffractionRD} 
\end{equation}
where we have assumed that $L$ is much larger than the coherence 
length $1/(2k_o - k_z)$. As expected, the emerging beam also has a 
Gaussian cross section, with a contributing volume determined by the 
size of the original beam and the coherence length of the 
configuration.

For the $x$ and $y$ SD components and the $y$ MG 
component the integrals are also standard. Taking the $x$ term 
as an example the result is 
\begin{equation}
	\begin{split}
		\sum_{\vec{r}_j} E_o (\vec{r}_j) \frac{ 
		\partial E_o ( \vec{r}_j) }{ 
		\partial x} & e^{i ( 2\vec{k}_o - \vec{k} ) \cdot 
\vec{r}_j} \\
		& = - \frac{ \pi N k_x W^2 E_o^2 }{4 (2 k_o - k_z ) } 
e^{- ( k_x^2 + k_y^2 ) W^2 / 8} . 
	\end{split}
\end{equation}
This is also a Gaussian beam, but with a nodal line passing through 
the center. This is the analytical representation of the two-lobed 
pattern reported by Figliozzi et al.~\cite{Figliozzi2005} for various 
configurations of SHG from amorphous material and spherical Si 
nanoinclusions.

\subsection{Net results\label{summary}}

We now combine the results of the above sections. The overall RD 
contribution is
\begin{equation}
	\begin{split}
		\vec{E}^{ff}_{RD} &= \frac{\pi \omega^3 q N W^2 E_o ^2 
}{30 r c^3 (2k_o - k_z) } \left[ \tilde{I} - \hat{k} \hat{k} \right] 
\cdot \\
		& \quad \quad \quad \left[ \hat{x} (\hat{k} \cdot 
\hat{x}) n_2 ( 3 C_{l}^2 + 4 C_{l}C_{t} + 8 C_{t}^2) \right. \\
		& \left. \quad \quad \quad \quad \quad + \left(\hat{y} 
(\hat{k} \cdot \hat{y}) n_2 + \hat{z} (\hat{k} \cdot \hat{z}) n_2 
\right) (C_{l} - C_{t})^2 \right] \\
		& \quad \quad \quad \times e^{-(k_x^2 + k_y^2) W^2 /8} 
e^{ikr - i2 \omega t}. 
	\end{split}
	\label{rdf} 
\end{equation}
The corresponding expressions for SD and MG are respectively
\begin{equation}
	\begin{split}
		\vec{E}^{ff}_{SD} & (\vec{r},t) = - \frac{\pi \omega 
^3 q N W^2 E_o ^2}{ 15 r c^3 (2k_o - k_z) } \left[ \tilde{I} - \hat{k} 
\hat{k} \right] \cdot \\
		& \left[ \hat{x} (\hat{k} \cdot \hat{x}) n_2 \left( 3 
C_{l} D_{l} + 2 C_{l} D_{t} + 2 C_{t} D_{l} + 8 C_{t} D_{t} \right) 
\right. \\
		& \left. \quad + \left( \hat{y} (\hat{k} \cdot 
\hat{y}) n_2 + \hat{z} n_1 \right) \left( C_{l} - C_{t} \right) \left( 
D_{l} - D_{t} \right) \right] \\
		& \times e^{-(k_x ^2 + k_y ^2) W^2)/8} e^{ikr - 
i2\omega t } ;
	\end{split}
	\label{sdf} 
\end{equation}
\begin{equation}
	\begin{split}
		& \vec{E}^{ff}_{MG} = \frac{ \pi \omega^3 q N W^2 
E_o^2}{3 r c^3 (2k_o - k_z)} \left[ \tilde{I} - \hat{k}\hat{k} \right] 
\cdot \\
		& \quad \quad \quad \quad \left[ \hat{y} (\hat{k} 
\cdot \hat{y}) n_2 + \hat{z} n_1 \right] \left( C_{t} D_{l} + 2 C_{l} 
D_{t} \right) \\
		& \quad \quad \quad \quad \quad \quad \times e^{-(k_x 
^2 + k_y ^2) W^2)/8} e^{ikr - i2\omega t } . 
	\end{split}
	\label{mgf} 
\end{equation}
Equations \eqref{rdf}, \eqref{sdf}, and \eqref{mgf} give the far 
fields from the retardation, spatial-dispersion, and magnetic 
contributions, respectively. Despite the appearance of assorted phase 
factors at different stages of the derivation, to the extent that the 
refractive indices are real all net contributions have the same phase 
to within a plus or minus sign. The RD contributions in the two 
directions perpendicular to that of the polarization of the incident 
beam are equal, as expected by symmetry. This is not the case for SD 
and MG, since SD involves gradients and $\vec{B}$ is an axial vector.

As noted in the Introduction, the linear response cannot be calculated 
by factoring as done above. A full self-consistent Ewald-Oseen 
treatment is necessary.

\section{Discussion\label{discussion}}

Although Eqs.~\eqref{rdf}, \eqref{sdf}, and \eqref{mgf} are complete, 
their general properties are not immediately evident. Hence we 
consider special cases. We also estimate conversion efficiency for 
fused silica, basing our calculations on several assumptions and the 
known LO properties of this material.

\subsection{Paraxial-ray approximation\label{parax}}

In the usual case of a highly collimated source beam of relatively 
small cross section, the emerging beam will also be initially 
relatively well localized but will diverge over a solid angle where 
the components essentially add in phase. Taking the diameter of the 
cross section W to be equal to at least a few wavelengths of the 
emerging beam, we make the paraxial-ray approximation, writing the 
observation direction for forward scattering as $\hat{k}$ = 
$\hat{x}\theta _x + \hat{y} \theta _y + \hat{z}$, where the 
beam-divergence (observation) angles $\theta_x$ and $\theta_y$ 
are first-order quantities. With this representation the various 
projection operations are easily evaluated and we find
\begin{equation}
	\begin{split}
		[(\tilde{I} - \hat{k}\hat{k}) &\cdot \hat{x}] (\hat{x} 
\cdot \hat{k}) = \hat{x} \theta_x ; \\
		[(\tilde{I} - \hat{k}\hat{k}) &\cdot \hat{y}] (\hat{y} 
\cdot \hat{k}) = \hat{y} \theta_y ; \\
		(\tilde{I} - \hat{k}\hat{k}) &\cdot \hat{z} = -\hat{x} 
\theta_x - \hat{y} \theta_y . 
	\end{split}
	\label{proj} 
\end{equation}
Considering also Eqs.~\eqref{rdf}--\eqref{mgf}, it is apparent that 
all contributions vanish in the forward direction and exhibit 
two-lobed patterns characteristic of gradient effects. Note that the 
$z$ component also contributes on the same first-order scale when the 
viewer is off-axis.  We shall use these equations in the following.

\subsection{Reduction to the isotropic case for large 
$\kappa_1$\label{red1}}

If the polarizable points are isotropic then $C_l = C_t = C$ and $D_l 
= D_t = D$. If we assume further that $\kappa_1 \gg 4m\omega^2$ then 
$C \approx D$.  For clarity we write
\begin{equation}
	g(\vec{k},r,t) = \frac{\pi \omega^2 N W^2 E_o^2 C^2}{8 r 
c^2}e^{-(k_x^2 + k_y^2) W^2 /8} 
e^{ikr - i2 \omega t},
\end{equation}
since this is a common factor for all cases discussed in the rest of 
Sec.~\ref{discussion}.  Then Eqs.~\eqref{rdf}--\eqref{mgf} reduce to
\begin{equation}
		\vec{E}^{ff}_{RD} = - 2 \hat{x} n_2 \theta_x 
\frac{g(\vec{k},r,t)} { (n_2 - n_1) } ;
	\label{isord}
\end{equation}
\begin{equation}
		\vec{E}^{ff}_{SD} = 4 \hat{x} n_2 \theta_x 
\frac{g(\vec{k},r,t)} { (n_2 - n_1) } ;
	\label{isosd} 
\end{equation}
\begin{equation}
		\vec{E}^{ff}_{MG} = 4 \left[ \hat{x} n_1 \theta_x - 
\hat{y} (n_2 -n_1) \theta_y \right]  \frac{g(\vec{k},r,t)} { (n_2 - 
n_1) };
	\label{isomg} 
\end{equation}
where we have used the fact that $k_z$ differs from $k$ only by terms 
of second order in $\theta$.  The net result is
\begin{equation}
	\vec{E}^{ff}_{Net} = 2 \left[ \hat{x} \left( 2 n_1 + n_2 
\right) \theta_x - \hat{y} 2 (n_2 -n_1) \theta_y \right]  
\frac{g(\vec{k},r,t)} { (n_2 - n_1) }.
	\label{isonet} 
\end{equation}
This limit applies to the SHG response of systems where the bond 
charge is strongly bound, for example organic materials and glasses. 
All three mechanisms generate polarization in the direction $\hat{x}$ 
of the applied field, and all have similar magnitudes, so none can be 
neglected. The fact that the RD term is important may not be at 
variance with the conclusion of Brudny et al.~\cite{Brudny2000} which 
pertains to a configuration where the anharmonic contribution does not 
vanish completely.  In particular, the RD contribution here is exactly 
half that of SD and with opposite sign, so the net effect of the RD/SD 
combination is to reduce the SD contribution by half. For $n_1 \approx 
n_2$ the magnitude of the forward-scattered field intensity clearly 
benefits significantly from a long coherence length.

Equation~\eqref{isomg} shows that two $\hat{y}$ contributions are 
present, but to the extent that $n_1 \approx n_2$ the overall term 
is small and can easily be overlooked since detection depends on 
intensity, not fields.  This near-cancellation is a result of the sign 
of the $z$ contribution in off-axis viewing.  The cancellation of the 
$y$ component is exact in Eq.~\eqref{rdf}, even in the general case 
where $C_l \ne C_t$.  A near-cancellation of the $y$ component also 
occurs in the general case for Eq.~\eqref{sdf}, although a second 
near-cancellation contributes if $C_l \approx C_t$ or $D_l \approx 
D_t$.  Thus if the $y$ component is analyzed quantitatively, the more 
general equations must be used. We conclude that the $z$ component is 
important in determining the properties of the emerging beam.  

\subsection{Metals\label{met}}

A second limit of the above is that corresponding to those metals 
for which the effective mass of the carriers is itself essentially 
isotropic. As a result of strong attenuation of optical signals, 
metals are usually measured in backscattering, where the results can 
be further complicated by surface 
reconstructions.~\cite{Tom1986ek,Rudnick1971rm,Driel1994rt} Although 
much attention has been focused on these surface 
contributions,\cite{Bloembergen1968lr,JHA1965,Driel1994rt,Murphy1989zl} 
we consider here only signals originating in the bulk.  In absorbing 
media the Green function retains its form, so the above development is 
still valid although the propagation vectors are now generally 
complex.  With no restoring force $\kappa_1$ = 0, so $C$ = $4D$. In 
the paraxial-ray approximation the RD contribution is unchanged, but 
the SD and MG terms are reduced by a factor of 4. For forward 
scattering the equations are 
\begin{equation}
		\vec{E}^{ff}_{RD,m} = - 2 \hat{x} n_2 \theta_x 
\frac{g(\vec{k},r,t)} { (n_2 - n_1) } ; 
	\label{metrd} 
\end{equation}
\begin{equation}
		\vec{E}^{ff}_{SD,m} = \hat{x}  n_2 \theta_x 
\frac{g(\vec{k},r,t)} { (n_2 - n_1) } ; 
	\label{metsd} 
\end{equation}
\begin{equation}
		\vec{E}^{ff}_{MG,m} = \left[ \hat{x} n_1 \theta_x  - 
\hat{y} (n_2 - n_1) \theta_y \right] \frac{g(\vec{k},r,t)} { (n_2 - 
n_1) } ; 
	\label{metmg} 
\end{equation}
\begin{equation}
		\vec{E}^{ff}_{Net,m} = - (n_2 - n_1) \left[ \hat{x} 
\theta_x  - \hat{y} \theta_y  \right] \frac{g(\vec{k},r,t)} { (n_2 - 
n_1) } . 
	\label{metnet} 
\end{equation}
The $x$ and $y$ components now have equal amplitudes, but 
to the extent that $n_1 \approx n_2$ the net result shows that the 
enhancement of the signal strength that results from the nearly 
singular denominator is cancelled.  Again, all three contributing 
mechanisms are important.  Thus the assumption that the SHG 
contribution from metals arises entirely from spatial dispersion and 
magnetic effects is not quite 
correct.~\cite{Rudnick1971rm,Driel1994rt}

SHG signals from surface reconstructions could be described in the 
above formalism by assigning suitable anisotropies to electrons in the 
surface region, although we do not do this here.

\subsection{Backscattering\label{back}}

For backscattering the major difference is the reduction of the 
correlation length and corresponding reduction in the radiated field, 
since for negative $k_z$ the two terms of $(2k_o - k_z)$ add instead 
of subtract.  As we shall show in Sec.~\ref{intensities}, this 
effectively eliminates any possibility of observing SHG from the bulk 
of amorphous materials.  The other effect is to reverse the sign of 
the result of the projection operation on $\hat{z}$.  When everything 
is taken into account, the paraxial-ray expressions for $\kappa_1 \gg 
4m\omega^2$ are
\begin{equation}
		\vec{E}^{ff}_{RD,b} = 2 \hat{x} n_2 \theta_x 
\frac{g(\vec{k},r,t)} { (n_1 + n_2) } ; 
	\label{backrd} 
\end{equation}
\begin{equation}
		\vec{E}^{ff}_{SD,b} =  - 4 \hat{x} n_2 \theta_x 
\hat{x} \frac{g(\vec{k},r,t)} { (n_1 + n_2) } ; 
	\label{backsd} 
\end{equation}
\begin{equation}
		\vec{E}^{ff}_{MG,b} = 4 [ \hat{x} n_1 \theta_x  + 
\hat{y} (n_1 + n_2) \theta_y ]  \frac{g(\vec{k},r,t)} { (n_1 + n_2) } 
;
	\label{backmg} 
\end{equation}
\begin{equation}
		\vec{E}^{ff}_{Net,b} =  2 [ \hat{x} (2 n_1 - n_2) 
\theta_x  + 2 \hat{y} (n_1 + n_2) \theta_y ]  \frac{g(\vec{k},r,t)} { 
(n_1 + n_2) } .
	\label{backnet} 
\end{equation}
While both polarizations are present, the dominant contribution in 
backscattering is that perpendicular to that of the driving field.  
Although an $x$ contribution is still generated, its strength is 
expected to be small compared to that polarized along $y$.

The expressions for isotropic metals are
\begin{equation}
		\vec{E}^{ff}_{RD,mb} = 2 \hat{x} n_2 \theta_x 
\frac{g(\vec{k},r,t)} { (n_1 + n_2) } ; 
	\label{metrdb} 
\end{equation}
\begin{equation}
		\vec{E}^{ff}_{SD,mb} = - \hat{x} n_2 \theta_x 
 \frac{g(\vec{k},r,t)} { (n_1 + n_2) } ; 
	\label{metsdb} 
\end{equation}
\begin{equation}
		\vec{E}^{ff}_{MG,mb} = [ \hat{x} n_1 \theta_x 
+ \hat{y} (n_1 + n_2) \theta_y  ] \frac{g(\vec{k},r,t)} { (n_1 + n_2) 
}; 
	\label{metmgb} 
\end{equation}
\begin{equation}
		\vec{E}^{ff}_{Net,mb} = (n_1 + n_2 )[\hat{x} \theta_x 
+ \hat{y} \theta_y  ] \frac{g(\vec{k},r,t)} { (n_1 + n_2) }. 
	\label{metnetb} 
\end{equation}

\subsection{Power and conversion efficiency\label{intensities}}

In many experiments what is determined is not the SHG intensity but 
the integrated SHG power.  To obtain an order-of-magnitude estimate we 
consider the net $x$-polarized component for forward scattering with 
$\kappa_1 \gg 4m\omega^2$ and with $C_l = C_t = C$ and $D_l = D_t = 
D$. The SHG intensity is given by 
\begin{equation}
	I_{SH} = \frac{cn_2}{8\pi} |\vec{E}^{ff}_{Net}|^2. 
\end{equation}
The SHG power is obtained by integrating this expression 
over a hemisphere of radius r. We are also interested in the 
conversion efficiency $\eta$, which we define as 
\begin{equation}
	\eta = \frac{P_{SH}} {(P_{inc})^2}, 
\end{equation}
where $P_{inc}$ is the power of the incident beam. Assuming that the 
incident beam is collimated, the evaluation of its power in terms 
of the beam properties is straightforward, and we obtain
\begin{equation}
	\begin{split}
		P_{inc} & = \int_{-\infty}^{\infty} dx \int_{- 
\infty}^{\infty} dy \frac{cn_1}{8\pi} | E_o |^2 e^{-2( x^2 + y^2 
)/W^2} \\
		& = \frac{cn_1}{16} W^2 |E_o |^2 . 
	\end{split}
\end{equation}

That for the emerging beam is more complicated. The first issue 
concerns angular dependences.  If the incident beam is reasonably well 
collimated and its diameter is equal to at least several SHG 
wavelengths, the SHG beam is also fairly well collimated.  Then a 
small-term expansion in $\theta$ is a good approximation. To show this 
we consider 
\begin{equation}
	e^{-(k_x^2 + k_y^2)/8W } = e^{-(k^2 sin^2 \theta )/8W } . 
\end{equation}
Taking $k = 2\pi n_2 /\lambda _{SH}$, $n_2 = 1.3$, $\lambda_{SH} = 400 
nm$, and an incident beam width of 5 $\mu m$, we have $k^2/8W \approx 
50$. Hence the small-term approximation $sin\theta \approx \theta$ is 
acceptable.  This also provides justification for our use of the 
paraxial-ray approximation in the previous sections. With these 
simplifications the area integral is straightforward and we find for 
$\hat{x}$ polarization 
\begin{equation}
	\begin{split}
		P_{SH,x} & = \frac{cn_2}{8\pi} \frac{\pi^2 
\omega^4 q^2 W^4 N^2 E_o^4 C^4}{16 c^4 ( n_2 - n_1)^2} ( 2 n_1 + n_2 
)^2  \\
		& \quad \quad\times \int_{0}^{2\pi} d\phi \int_0 
^{\infty} \theta d\theta (\theta^2 cos^2\phi) e^{-k^2 \theta^2 
W^2/4}\\
		& = \frac{\pi^2 c q^2 N^2 E_o^4 C^2 }{64  
n_2^3 ( n_2 - n_1 )^2 } ( 2 n_1 + n_2 )^2 . 
	\end{split}
\end{equation}
Combining the above expressions we find the corresponding conversion 
efficiency to be: 
\begin{equation}
	\eta_x = \frac{4 \pi^2 q^2 N^2 C^4 (2n_1 + n_2)^2 }{c W^4 
n_1^2 n_2^3 (  n_2 - n_1)^2}. 
\end{equation}
The efficiency decreases as the fourth power of the diameter of the 
incident beam. This is in contrast to the intensity, which decreases 
as $1/W^6$.

From the definition of $C_{l}$ we have 
\begin{equation}
	\vec{p} =  \alpha \vec{E}_{loc} = q \Delta \vec{r} = q C_l
\vec{E}_{loc} ,
\end{equation}
where $E_{loc}$ is the field at the charge site and $\alpha$ is the 
linear polarizability.  Then we can write $C = \alpha/q$.
We can connect $\alpha$ to the dielectric 
function $\epsilon_1 = n_1^2$ and bond density $N$
of the material by the Clausius-Mossotti relation 
\begin{equation}
	\frac{4\pi}{3} N \alpha = \frac{\epsilon_1 - 1}{\epsilon_1 + 2} . 
\end{equation}
Then
\begin{equation}
	\eta_x = \frac{81 ( 2 n_1 + n_2 )^2 }{64 \pi^2 c q^2 N^2 W^4 
n_1^2 n_2^3 ( n_2 - n_1)^2} \left(\frac{\epsilon_1 - 1}{\epsilon_1 + 
2}\right)^4. 
\end{equation}
Using a driving wavelength $\lambda = 800 \, nm$, dielectric functions 
of quartz of 2.112 and 2.161 at 800 and 400 $nm$, respectively, a bond 
density of $1.06 \times 10^{23} \, cm^{-3}$,~\cite{1977qy} 
and a Gaussian beam of characteristic dimension $W = 10 
\mu m$ we find $\eta_x = 1.4 \times 10^{-18} \, watt^{-1}$. Thus 
1 $watt$ input power at 800 $nm$ is expected to generate about 3 SHG 
photons/sec. If $W$ is reduced to $1 \mu m$, the output would increase 
to about $10^4$ SHG photons/sec. These results appear to be consistent 
with experiment,~\cite{Figliozzi2005} where few if any photons were 
seen emerging from the glass substrate. 

\section{Conclusions\label{sec-conclusions}}

We have developed an anisotropic bond model (ABM) that describes SHG 
on the atomic scale, uses physically meaningful parameters, and 
includes all contributing mechanisms, thereby providing a more 
complete understanding of the physics of SHG than previously 
available. In disordered materials the anharmonic restoring force 
acting on the bond charge does not contribute to the overall SHG 
signal, which instead arises from a combination of LO and 
beam-geometry effects and therefore provides limited new information 
about the material.  For a Gaussian driving beam we obtain analytic 
expressions that give the phase, amplitude, and spatial distribution 
of the SHG radiation field for each of the remaining contributing 
mechanisms:  retardation (RD), spatial dispersion (SD), and 
magnetic-field (MG) effects.  All have the same order of magnitude, so 
any complete description must consider each.  The expressions are 
reduced to simpler forms for both forward- and backscattering 
configurations in two isotropic limits, the first where the linear 
restoring force dominates, as in glasses, and the second where it is 
absent, as in metals.  We estimate the conversion efficiency for 
forward scattering in fused quartz.  Predictions appear to be in 
agreement with observations, where available.~\cite{Figliozzi2005} 
Specific additional predictions allow critical tests of these results. 

With the basic physics established, we can now consider more 
complicated configurations, including nanospherical inclusions in 
glass and the reported SHG enhancement with crossed-beam, 
crossed-polarization driving fields.~\cite{Figliozzi2005} 
The results presented here are also expected to be useful for 
analyzing SHG data of liquids and biological materials.


\end{document}